\documentclass{elsart3}

\usepackage{natbib}

\usepackage{epsfig}

\usepackage{amssymb}

\begin{document}

\begin{frontmatter}



\title{Contact binary stars of the W~UMa type as distance tracers}


\author{Slavek M. Rucinski}
\ead{rucinski@astro.utoronto.ca}
\address{David Dunlap Observatory, University of Toronto\\
P.O.Box 360, Richmond Hill, Ontario, Canada L4C~4Y6}

\begin{abstract}

Contact binaries can be used for distance determinations of
stellar systems. They are easy to discover and identify and are
very abundant among solar-type stars, particularly
for $M_V > +3$. The period--luminosity--colour (PLC) relations
have similar properties to those for pulsating stars and can currently
predict individual values of $M_V$ to about $\pm 0.25$ mag. 

\end{abstract}


\end{frontmatter}

\section{Introduction}
\label{intro}

Contact binary stars of the W~UMa-type (also known as
EW) are unique objects \citep{ruc93,web03}. 
The luminosity, produced almost exclusively 
in the more massive component is efficiently 
distributed (by poorly understood processes)
through the common envelope  
so that the surface brightness is practically
identical over the whole visible surface of the binary
(the gravity variations over the surface introduce a very 
minor modification). Thus, the effective temperature is
everywhere -- over the whole common surface --  
the same, in spite of usually very different masses hiding
inside the common envelope. Mass ratios are known to
span the whole wide range, from almost unity (V753~Mon, 
$q = 0.97$, \citealp{ddo3}), to very small values,
as small as $q=0.066$ (SX~Crv, \citealp{ddo5}).
The more common  -- but more difficult to detect -- 
small mass-ratio systems are the best illustration to 
why the contact binaries are so unique. Simplifying: 
the primary component provides the luminosity, 
while both components provide the radiating area. 
Thus, it is practical to consider the primary mass, 
$M_1$, and the mass ratio, $q = M_2/M_1 \le 1$, as the two
main parameters. The third parameter, the orbital period, 
$P$, is related to the amount of the angular momentum
in the system. The range of the primary masses is moderate and
corresponds to Main Sequence spectral types from early
A to early K and roughly maps into the orbital-period range
of about 1.5 days to 0.22 days. Since the period distribution
appears to continue the $1/P$ shape (as for other binaries) 
to about 0.45 days, and shows a mild
bending down between 0.45 and 0.25 days, 
the most common are short period systems with periods
within 0.25 -- 0.5 days, down to the
sharp and currently unexplained cutoff at 0.22 days
\citep{ruc92,ste01}.

The contact binaries are known to exist only within the 
Main Sequence. Somewhat similar light curves 
of some giant or other more evolved
binaries can be explained by semi-detached 
binaries or spotted stars. Very rare early-type contact
systems on the upper Main Sequence, including a few currently known
O-type contact systems, do not appear to be very
different from the more typical solar-type W~UMa binaries, but 
will not be considered here.

With their special properties, the contact binaries of the W~UMa type
form a distinct group of objects which are very easy to 
detect and identify due to: (1)~rather large
amplitudes of light variations reaching one magnitude, (2)~short
orbital periods, so that limited--duration monitoring
is sufficient. Massive, systematic searches of the sky for 
stellar variability, such as the micro-lensing projects, 
have led to many detections of EW systems; 
e.g.\ in the first catalogue of OGLE-I,
as much as two thirds of the one thousand 
detected eclipsing binaries were EW systems
\citep{ruc97}. Many EW systems 
have been detected during the galactic cluster
searches, both in open and in globular clusters (see
\citealp{ruc98,ruc00} for the references to the original works).

\section{The $P \sqrt{\rho}$ relation for contact binaries}
\label{Q-rel}

Contact binaries are not as convenient as detached binaries 
for distance determinations. Although described by
fewer parameters than detached binaries (one potential in
place of two independent radii), they have
more complex geometry. As pointed by Dr.\ R.\ E. Wilson
during this conference, a single detached, eclipsing, double-line
spectroscopic binary can provide a distance determination 
without any calibration, basically using the
``first-principles'' approach. In this respect, the contact 
binaries are more like pulsating stars and -- 
with still many uncertainties about their structure and adherence
to the strict Roche model -- do require empirical luminosity
calibrations. 

The rationale for the existence of a period -- luminosity
relation is based in the strong geometric constraints
imposed by the common equipotential envelope which permit
to consider an equivalent of the $Q=P \sqrt{\rho}$ relation
for contact binary systems. To obtain this, one combines 
the total volume of the contact configuration, described
by the Roche common equipotential, with the
Kepler's law, $a^3/P^2 = G (M_1+M_2)$. The volume of
the binary system, $V$, van be written in non-dimensional
way as: $v = V/a^3$. Such a dimensionless
value, $v = v(q,F)$, does depend slightly
on the mass ratio $q$ and on the degree of contact, 
$0 \le F \le 1$ ($F=0$ for the
inner common equipotential surface with one point of contact
at the Lagrangian point $L_1$; $F=1$ for the outer
surface opening to the space at $L_2$). Tables of 
$v(q,F)$ have been calculated by \citet{moch84,moch85}.
It is observed that most of contact binaries have
$0 \le F \le 0.2$; however, this is a model-dependent
result \citep{ruc93}. 

\begin{figure}[!t]
\label{figQ}
\begin{center}
\includegraphics[scale=0.48]{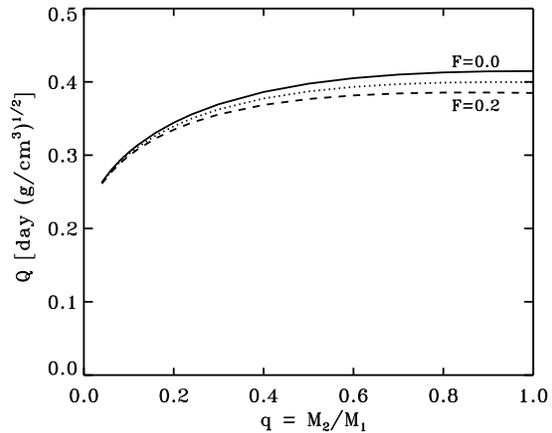}
\end{center}
\caption{The ``pulsation constant'' for contact binary stars
as a function of the mass ratio $q$ for $F=0, 0.1, 0.2$.}
\end{figure}

By defining the mean density of the whole configuration,
$\rho = (M_1 + M_2)/V$, then eliminating $a^3$, and by
expressing the quantities in familiar units, one obtains
the ``pulsation equation'',
$P \sqrt{\rho} = Q(q,F) = \sqrt{0.079/v(q,F)}$, with
$P$ in days and $\rho$ in g/cm$^3$ (Figure~\ref{figQ}). 
The similarity of contact binaries to pulsating stars is not
accidental as the underlying physical time scale, of the
pulsation or of the orbital revolution, is the same
familiar dynamical time scale.

\section{The basis of the luminosity calibration}
\label{basis}

One attempts to derives 
a luminosity calibration which would not depend on any 
spectroscopic data. Instead of the size of the orbit, $a$, 
we want to use the orbital period, $P$. 
Similarly as for the non-dimensional volume in the
previous section,
one can define the non-dimensional surface of the contact
binary, $s$, so that the total area, $S$ can be written
as $S = s \, a^2$. Again, $s = s(q,F)$, but this dependence is
weak (the tables are in \citealp{moch84,moch85}).
The luminosity of the system, can be written quite generally,
as: $L \propto S \,T_{\rm eff}^4 \propto s \,a^2 \,T_{\rm eff}^4$. 
We remove the unknown $a$ using the Kepler's 
law and, in its place, introduce the orbital
period, $P$, which is determinable with an accuracy by many orders
of magnitude higher than any other quantity. One can rewrite
all these as one equation for the absolute magnitude $M_V$:
\begin{eqnarray*}
M_V & = & -10 \log T_{\rm eff} + B.C.(T_{\rm eff}) - 10/3 \log P - \\ 
    &   & -5/3 \log M_1 - 5/3 \log (1+q) - \\
    &   & -2.5 \log s(q,F) + {\rm const}
\end{eqnarray*}
Note that the two first terms contain the $T_{\rm eff}$
dependence, then there is the ``infinitely well-known''
$P$--term and then there are terms dependent on $M_1$,
$q$ and $F$. As always, when photometric data are available,
$T_{\rm eff}$ can be estimated from the
colour index or from the spectral type, but the last terms
are not known and require spectroscopic data. 
We ``sweep them under the rug'' at this point 
to utilize the simplest possible calibration equation:
$$M_V = C_1 * \mathit{colour} + C_2 \log P + C_3$$
In this period--luminosity--colour relation (PLC), 
$\mathit{colour}$ is any of the available
colour indices such as $B-V$ or $V-I$; 
for each choice of the index, another set
of the coefficients $C_i$ must be determined. Lumping together
all $M_1$, $q$ and $F$ dependencies is not as artificial as
it may seem as several hidden correlations exists between
these parameters; some of these correlations are in fact 
of substantial importance for the
theory of the structure of these binaries \citep{ruc93}.

\section{Existing PLC calibrations}
\label{cals}

After a very early attempt \citep{ruc74}, the two main works dealing
with the subject were \citet{ruc94} and \citet{rd97}(=RD97), 
see also \citet{ruc96}. Several
small improvements were discussed later in the context
of the massive variable star searches, e.g. \citet{ruc98,ruc00}.
While \citet{ruc94,ruc98} were based mostly on the open cluster
data (EW systems appear at a relative frequency of some 1/500
-- 1/1000 in old open clusters), RD97 was based on the Hipparcos
data. The Hipparcos database remains the best source and has
recently been analyzed in great detail for the complete sample
of systems with $V < 7.5$ mag \citep{ruc02}.
The currently best calibrations utilizing de-reddened $B-V$ and
$V-I$ colour indices are:
\begin{eqnarray*}
M_V & = & -4.44\, \log P + 3.02\, (B-V)_0 +0.12\\
M_V & = & -4.43\, \log P + 3.63\, (V-I)_0 -0.31
\end{eqnarray*}

\begin{figure}[!t]
\begin{center}
\includegraphics[scale=0.50]{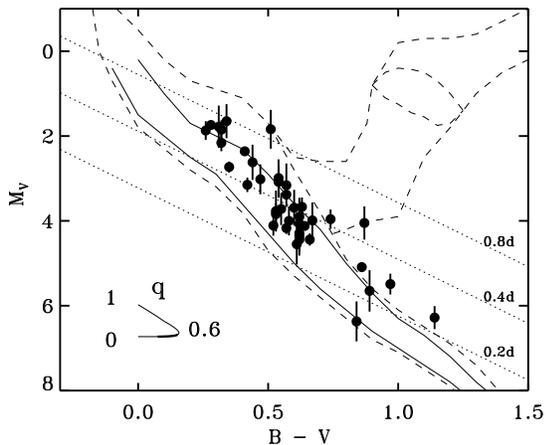}
\end{center}
\caption{The colour -- magnitude diagram for contact binaries
with the Hipparcos parallaxes. The $M_V$ errors are estimated
from the parallax errors.}
\label{figCMD}
\end{figure}

The colour--magnitude diagram (CMD) for the RD97 sample of contact
binaries with Hipparcos parallaxes is shown in Figure~\ref{figCMD}.
The thin continuous and broken contours delineate the density
of stars in the general Hipparcos database; one can see the 
Main Sequence, the Red Giant branch and the Red Clump. 
Each binary system
appears as one point corresponding to the combined brightness
of both components at light
maximum; this is in concordance with the fact
that there is just one (common) radiating area which radiates
the energy produced (mostly) in one, more massive star. 
The expected dependence on the
mass ratio is shown in the left corner of the figure. For $q=1$
two identical components provide equal amounts of luminosity so that
the shift is upwards by 0.75 mag; for $q<1$ the large radiating 
area, together with the diminished luminosity, result in lowering
of the effective temperature. For the most typical cases 
with $q$ around 0.3 to 0.5, there is almost no change of the
luminosity relative to the single star case, but the reddening
can reach up to 0.17 in $B-V$. As we can see in the figure,
the combined shift to the right and possibly evolution within
the Main Sequence have resulted in the EW systems following the
upper border of the MS for single stars.

The CMD in Fig. \ref{figCMD} contains also the
lines of equal orbital period, shown as slanted dotted lines. One can
directly see that correlation between the colour and the period
exists in that bluer, brighter systems have longer orbital periods.
The ``period--colour'' relation was discovered by \citet{egg67}.
On one hand, it complicates the matters by introducing an
internal correlation between the quantities in the $M_V$ calibration
so that they are not really ``orthogonal''. On the other hand,
it sometimes helps in weeding out variable stars which are of 
entirely different
type in that the data points cannot appear above and to the left
of a short-period blue envelope; they can only shift down to redder
colour indices and to the right due to the evolution \citep{ruc02}.

\section{Uncertainties of the calibrations}
\label{uncert}

There are several sources of uncertainties in the calibrations.
The trigonometric parallaxes from the Hipparcos mission introduce
errors typically $<0.25$ mag, but for a few systems in RD97 as large
as 0.5 mag. While such large errors can be accounted for by an
appropriate weighting, they lower the quality of the calibration.
In addition,
some of the binaries are unrecognized triple systems with an
associated offset in brightness, loss in accuracy and 
even entirely false data for the parallaxes. 
As for all calibrations involving
de-reddened colour indices, the $M_V$ estimates will depend on
how well this procedure is actually done. The simplification of
the calibration to the $\mathit{colour}$ and $\log P$
dependence, as described in Section~\ref{basis},
 also worsens the fit, but -- by far -- the main
source of errors is in the photospheric spots on individual binaries.
These binaries are very active and almost always show spots. A
brightness calibration must simply assume some sort of an average
for $M_V$.

When a simple mean weighted error is considered, then the deviations
in RD97 are characterized by $\sigma M_V \simeq 0.35$ mag. This is
however a pessimistic 
estimate because the Monte--Carlo simulations 
indicate that within the main range of the applicability, the
typical errors are $\sigma M_V \simeq 0.25$ mag (RD97).

\section{Metallicity dependence}
\label{FeH}

As with any calibration for pulsating stars, one must establish
the metallicity dependence of the PLC relation. At first,
when limited material was available, it seemed that such a dependence
did exist \citet{ruc95}, but -- when much more extensive data for
many globular clusters were added -- a metallicity term 
$\propto [Fe/H]$ became no longer needed \citep{ruc00}.

\begin{figure}[!t]
\begin{center}
\includegraphics[scale=0.50]{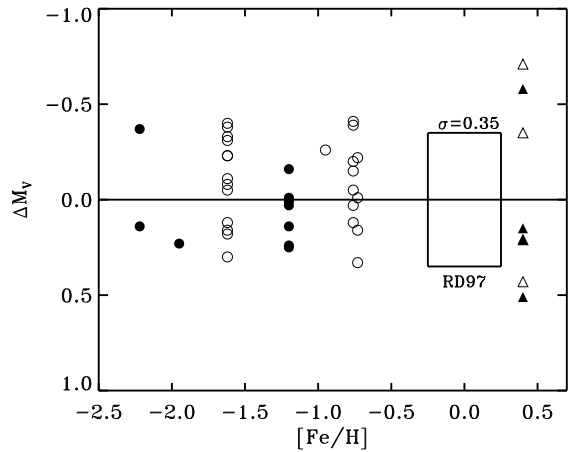}
\end{center}
\caption{Deviations from the $M_V$ calibrations 
for contact systems in several globular clusters. The box
labeled RD97 gives the schematic range of the parameters for
the Hipparcos sample. Closed and open symbols are used for
deviations from calibrations in $B-V$ and $V-I$, respectively. 
The data for $[Fe/H] > 0$ are for NGC~6791 which is an
open cluster.}
\label{fig_feh}
\end{figure}

\begin{figure*}[!t]
\begin{center}
\includegraphics[scale=0.88]{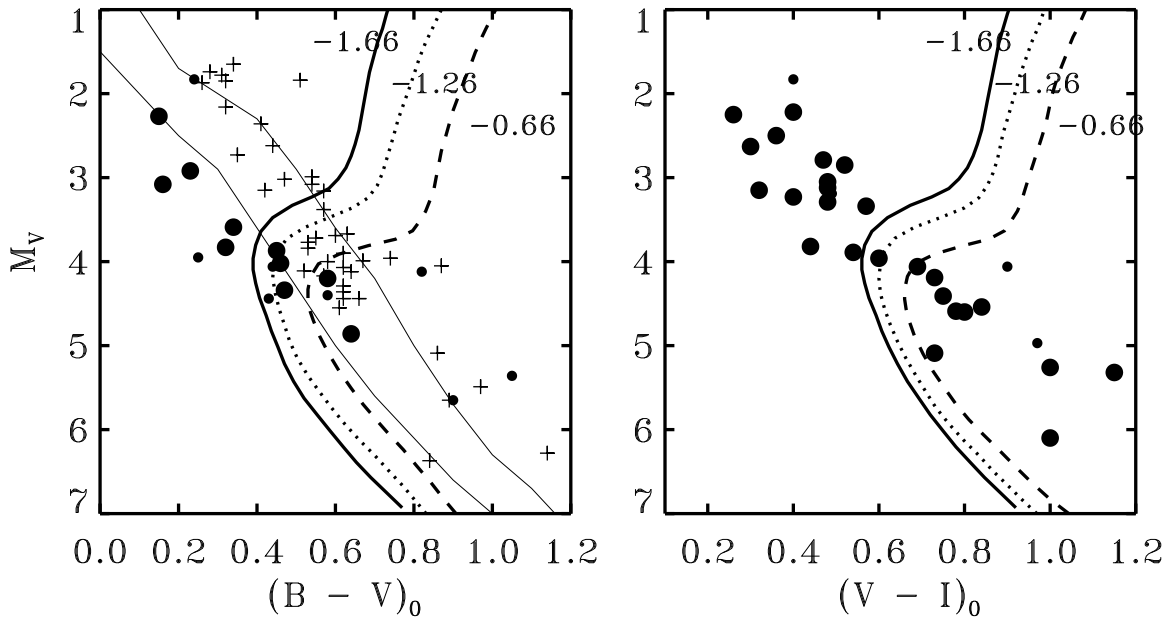}
\end{center}
\caption{Colour -- magnitude diagrams for contact
binaries in globular clusters. The 
isochrones for 3 values of $[Fe/H]$ are superimposed.
The left panel for $B-V$ shows also the data for the
solar neighbourhood, Hipparcos sample (crosses and
thin contours).}
\label{fig_gc}
\end{figure*}

The deviations from the $M_V$ calibrations, given in
Section~\ref{cals}, are shown as the
function of $[Fe/H]$ in Fig. \ref{fig_feh}. They are plotted for the
whole wide range of accessible metallicities ranging within
$-2.2 < [Fe/H] < +0.3$. The negative values of $[Fe/H]$ 
are provided by globular clusters of various ages, 
while the only positive metallicity is for the
rich, old, open cluster NGC~6791 \citep{WJ03}. There is no discernible 
trend in the figure so that the solar-neighbourhood calibrations
work well for contact systems of different metallicities, at
least at the level of uncertainty shown in the figure
($\sigma = 0.28$ and 0.27 for calibrations based on 
the $B-V$ (filled circles) and $V-I$ (open circles), 
respectively). These estimates of
the uncertainties may be exaggerated, because many additional
factors tend to increase the scatter; these could be errors
in the assumed reddening and absorption corrections, the 
large measurement errors for faint 
stars in distant, crowded clusters and the
uncertainties of the assumed distance moduli, $m-M$. 

In spite of what has been said above about lack of an
explicit dependence of $M_V$ on $[Fe/H]$, the contact binaries
with other metallicity content are different than the disk
population ones; it is only that at the current level of 
accuracy there exists no need to introduce the
metallicity corrections in $M_V$. Figure~\ref{fig_gc}
shows the CMD for systems in globular clusters
with available $B-V$ or $V-I$ \citep{ruc00}. 
The $B-V$ panel shows also, as crosses, the same Hipparcos 
(RD97) systems as in the figure in Section~\ref{cals}.
It is clear that the data points for globular-cluster binaries
are shifted left and below the disk--population Main
Sequence: the Population~II contact binaries are bluer
and smaller i.e.\ have shorter orbital periods. This latter
property is clearly seen in the period distribution for
globular cluster systems in \citet{ruc00}. Thus, the shift
in the CMD is mostly horizontal in $M_V$. This is
a preliminary result which must be checked again as the
data become more accurate.

\section{Misclassification and spatial density}
\label{spatial}

There are two final questions to answer: How many
EW systems can one expect relative to other stars? Is
there any danger of misclassification and taking
other variable stars for contact binaries or vice versa?

The only information at present relates to our Galaxy.
The EW binaries are very common in the Disk Population.
The estimate based on the OGLE-I pencil--beam search
volume gave the density as high as 1/130 among late-A to
early-K MS stars, a number consistent with some
old open clusters \citep{ruc98}. However, this high
estimate is not confirmed by a rigorous analysis of the
Hipparcos, solar-neighbourhood 
sample limited to stars brighter than
7.5 magnitude \citep{ruc02}; this analysis suggests the relative
density of about 1/500 (this corresponds
to the spatial density of $1.0 \times 10^{-5}$~pc$^{-3}$). 
It is at present not clear,
if the high density obtained from the OGLE-I sample
was due to the improperly accounted image blending or to a
genuine increase of the contact-binary density in the
central parts of the Thick Disk of our Galaxy.

The local volume around the Sun to 20 pc contains
about 300 Main Sequence stars of relevant spectral types
\citep{wie83}.
In this volume, we know only one EW system (44~Boo, also
known as i~Boo). This simple estimate agrees with the
lower number of the relative density of occurrence of the
EW systems. It is possible that the main reason for
the discrepancy in the density estimates is the mean
population age. The fractional density definitely increases
over time. It may be $\le 1/1000$ at 1~Gy, about 
1/500 at 5~Gyr and may reach 1/100 for ages above that.
Note that the fractional density among blue stragglers
in globular clusters appears to be as high as 1/45 \citep{ruc00}.

The variable stars which can most likely be taken for
contact binaries are short-period pulsating stars in the
$\delta$~Sct instability strip. 
Dr. G. Handler (private communication) estimates that about
1/3 of stars within the $\delta$~Sct instability strip are
actually pulsating. This is confirmed by the statistics
for the solar neighbourhood: Within the appropriate range
of the luminosity and colour, there are 32 stars within 
$<20$ pc and 8 are known to pulsate so that the factor 1/3
is confirmed within the Poisson fluctuation limit. 

The horizontal branch, Population~II, short-period
pulsators, the RR~Lyr variables, are very infrequent 
in the solar neighbourhood (the spatial density 
$6.2 \times 10^{-9}$~pc$^{-3}$, hence 0.0002 within 20 pc around
the Sun, \citealp{sun91}). Also infrequent are the 
MS Population~II Main Sequence pulsators, the SX~Phe stars;
their density can be roughly estimated as about 
1/100 of the Population~I $\delta$~Sct stars. 
All these variables occur within narrow ranges
of luminosities and spectral types, whereas contact binaries
can consist of stars of any spectral types throughout the
whole Main Sequence down to about K2 -- K4 (it is 
not clear why they do not exist below this limit). 

\begin{figure*}[!t]
\begin{center}
\includegraphics[scale=0.88]{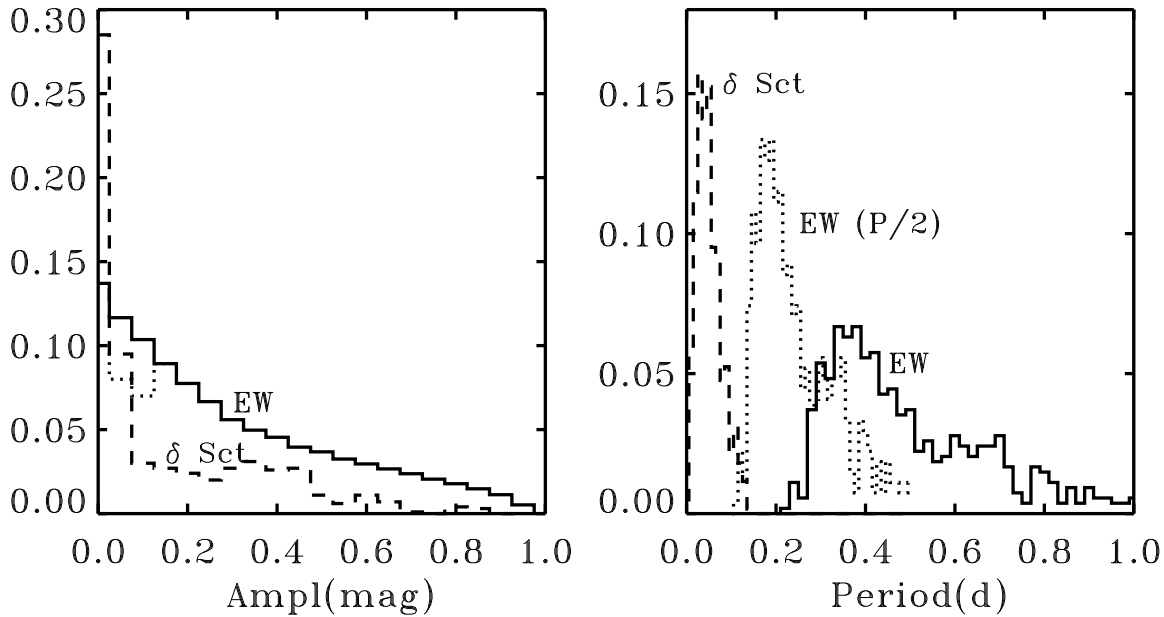}
\end{center}
\caption{The left panel shows the amplitude distribution
for contact binaries (EW, continuous line) 
and for $\delta$~Sct stars. The right panel shows
the period distributions for these two types of
variables. The dotted histogram shows the expected
distribution for contact binaries under assumption that
{\it all have been assigned incorrect periods\/} equal
to one half of the true orbital periods.}
\label{fig_dist}
\end{figure*}

Taking the estimates given above together, one finds that
there are about $8 \times$ more $\delta$~Sct stars in the
solar vicinity than EW binaries. However, there should be
no great difficulty in distinguishing them apart as their
light-variation amplitude and period distributions are quite
different. $\delta$~Sct tend to have very small amplitudes,
$<0.1$ mag, although large-amplitude variables do occur,
see Figure~\ref{fig_dist} based on \citet{RB01}.
But the most distinguishing is the period distribution:
The $\delta$~Sct stars have very short periods, $<0.15$ days
(broken line).
The contact binaries have longer periods starting just above
0.2 days. Even if one assigns a wrong period to a EW system
equal to $P/2$ (dotted), 
this period in most cases will turn out to
be still too long for a typical $\delta$~Sct star.

\section{Conclusions}
\label{concl}

W~UMa-type (EW) contact binaries are very easy to find in massive
stellar variability programs, even in short-duration ones, thanks
to the short orbital periods and large amplitudes of photometric
variations. The $M_V$ calibrations utilizing de-reddened colour
indices $B-V$ or $V-I$ and 
the periods can predict individual values to about
$\pm 0.25$ mag, so that about a thousand systems are needed
to reduce the group uncertainty to the level of $\pm 0.01$.
Such large numbers will be discovered in the nearby galaxies once
the surveys pass the threshold of $M_V \simeq 3 - 5$ which,
for the Local Group typically 
corresponds to $V > 23 - 25$. The current calibrations
involve de-reddened colour indices and thus remain sensitive to
the reddening corrections. The widespread availability of the 
$K$-band data suggests development of
calibrations based on the $V-K$ or
$I-K$ indices; this is a matter of the current work.
The existing $M_V$ calibrations do not seem to require
a correction for metallicity, but existence of a small $[Fe/H]$
term is not excluded.

\medskip

Thanks are due to Dr.\ G.\ Handler for illuminating e-mail discussion
on the density of the Main Sequence pulsating stars.




\end{document}